\documentclass[prl,12pt,citeautoscript,reprint]{revtex4-1}
\usepackage{graphicx}
\usepackage{epsfig}
\usepackage{amsmath}
\usepackage{graphicx}
\usepackage{dcolumn}
\usepackage{bm}
\usepackage{amssymb}
\usepackage{amsmath}
\usepackage{epsf}
\usepackage{subfigure}
\usepackage{epstopdf}
\usepackage{color}
\usepackage{caption}
\usepackage{subeqnarray}
\usepackage{mathrsfs}
\usepackage[utf8]{inputenc}
\DeclareUnicodeCharacter{0177}{\^y}
\usepackage[colorlinks=true, pdfstartview=FitV, linkcolor=red, citecolor=blue, urlcolor=blue]{hyperref}
\newcommand{\be}{\begin{equation}}
\newcommand{\ee}{\end{equation}} 
\newcommand{\ben}{\begin{eqnarray}} 
\newcommand{\een}{\end{eqnarray}} 
\begin{document}

\title{Time-Series Analysis of Internet Congestion Data}
\author{Anishi Mehta (ID: 201401439)}

\affiliation{Advisor: Dr. Amogh Dhamdhere, CAIDA, University of California, San Diego}

\begin{abstract}
There has been a lot of discussion on Net Neutrality and policies that various network service providers and distributors adopt, at times leading to greater network congestion and thus more debates. The aim of this project is to use congestion traffic data to look at correlations between different service providers and message requests for different AS (Autonomous Systems). The RTT (Round Trip Time) from the time-series data of these messages is evaluated, to provide conclusive results of favoring certain websites over others. Lastly, this project attempts time-series prediction to forecast what a the time series will look like a few hours or days from now given the history of what it looked like before.\\
\\
\textit{\textbf{Glossary} - Congestion traffic, inter-domain congestion, time-series analysis, RTT, net-neutrality}
\end{abstract}

\maketitle

\section{Introduction}
There is a rapid increase in the amount of internet traffic that demands a higher bandwidth. There is also a greater concentration of content among a smaller group of providers and distributors. All these lead to greater network congestion. This leads to a reduction in quality of services received by the user as well as disputes between service and content providers regarding the charges for increase in link capacity\cite{ref5}. Looking at the congestion patterns in the network could lead to mutual consensus regarding these disputes.\\
\\
CAIDA (Center for Applied Internet Data Analysis) is a leading organization that collects and analyzes real-time internet traffic data. They provide insights into internet infrastructure and its evolving nature by analyzing data to contribute to the fields of Internet science and communications public policies. \\
\\
CAIDA has access to Border Routers for Autonomous Systems (AS) across the world, and the time series data of the congestion traffic is stored in a database using InfluxDB. This database stores data from 2006 and is updated every five minutes. CAIDA uses this time series data to gain inferences regarding congestion rates for specific paths, Internet Service Providers (ISP) and destinations. \\
\\
This project uses the Pearson Correlation Coefficient to determine correlations in the measured Round Trip Time (RTT) for different packages over two weeks time on different ISPs and for different destinations. A high correlation in congestion rates between two time series, while one had low correlations with other time series, implies that the congestion is unnaturally higher on that link. This could be an indication of the following things: (a) The link used by the packet has a low capacity due to infrastructural issues (b) Packets are being intentionally dropped due to disputes between service providers and content providers.\\
\\
This project also attempts to predict future time-series based on the patterns observed from historical data. The bases of this is to predict RTT for a new packet sent to a destination IP via particular AS link on a given ISP. This is done using the Holt-Winters method for time-series prediction which will be explained later.

\section{Congestion Data}

Network congestion is the phenomenon of reduction in quality of service that occurs when a network path carries more data than its capacity. This may lead to a loss in packets and queuing delay in many cases. In practical terms, a high network congestion results in a web page loading incompletely or slowly, or a video buffering for too long on live video streaming websites such as YouTube and Netflix. \\
\\
A previous study done at CAIDA showed that among the top 10 US ISPs, 82\% of traces were a single AS hop \cite{ref5}. Thus, each interconnecting link consists of a near-side and far-side border router. Each time series records the RTT for both routers and a separate analysis needs to be done for both.

\begin{center}
    \centering
    \includegraphics[scale=0.22]{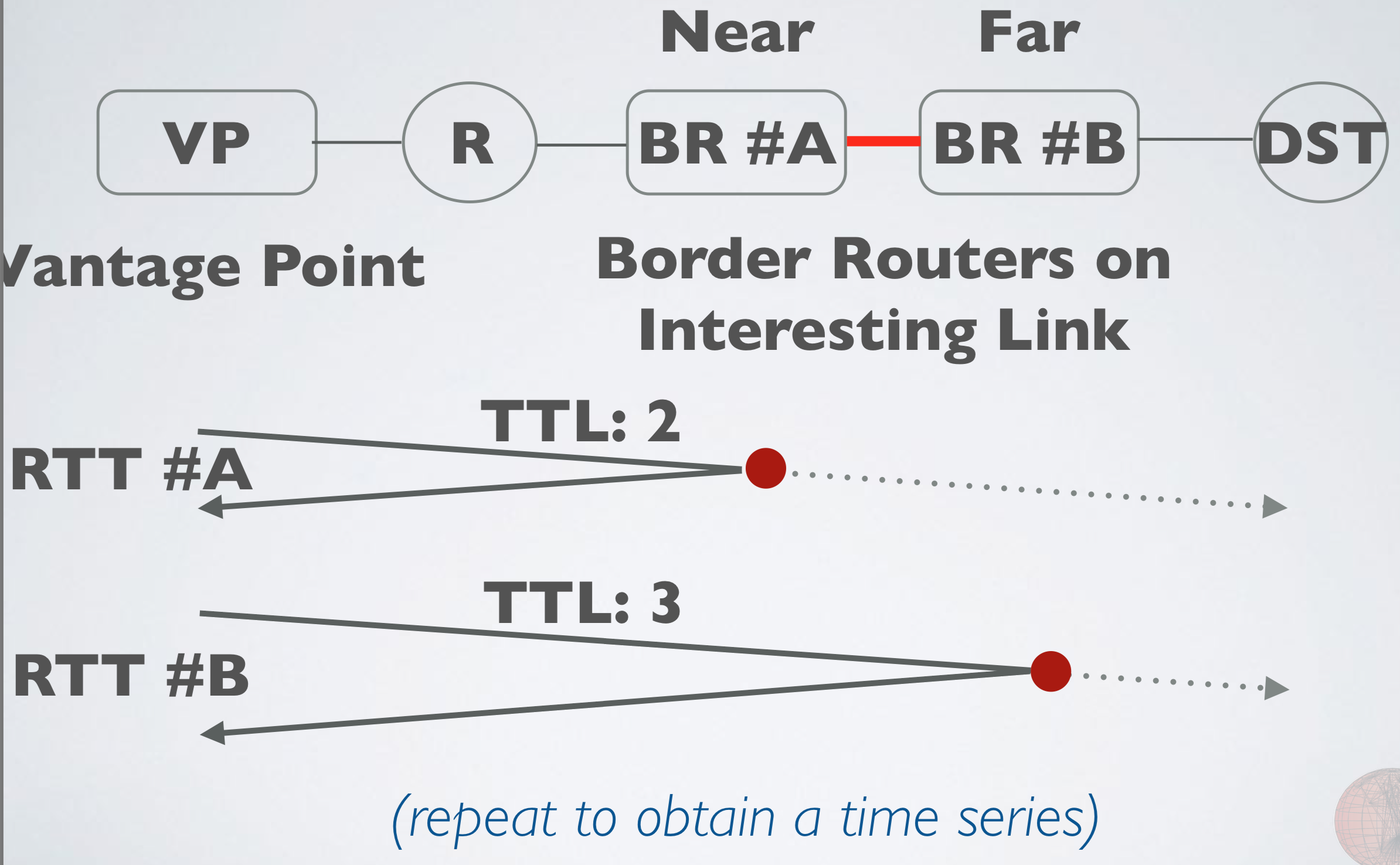}
    \captionof{figure}{Depiction of time series generation for network congestion data. The probe is sent to near-side router and reports the RTT on its return. Similarly, for far-side.. This is repeated periodically to generate a time series.}
    \label{fig1}
\end{center}

\begin{center}
    \centering
    \includegraphics[scale=0.21]{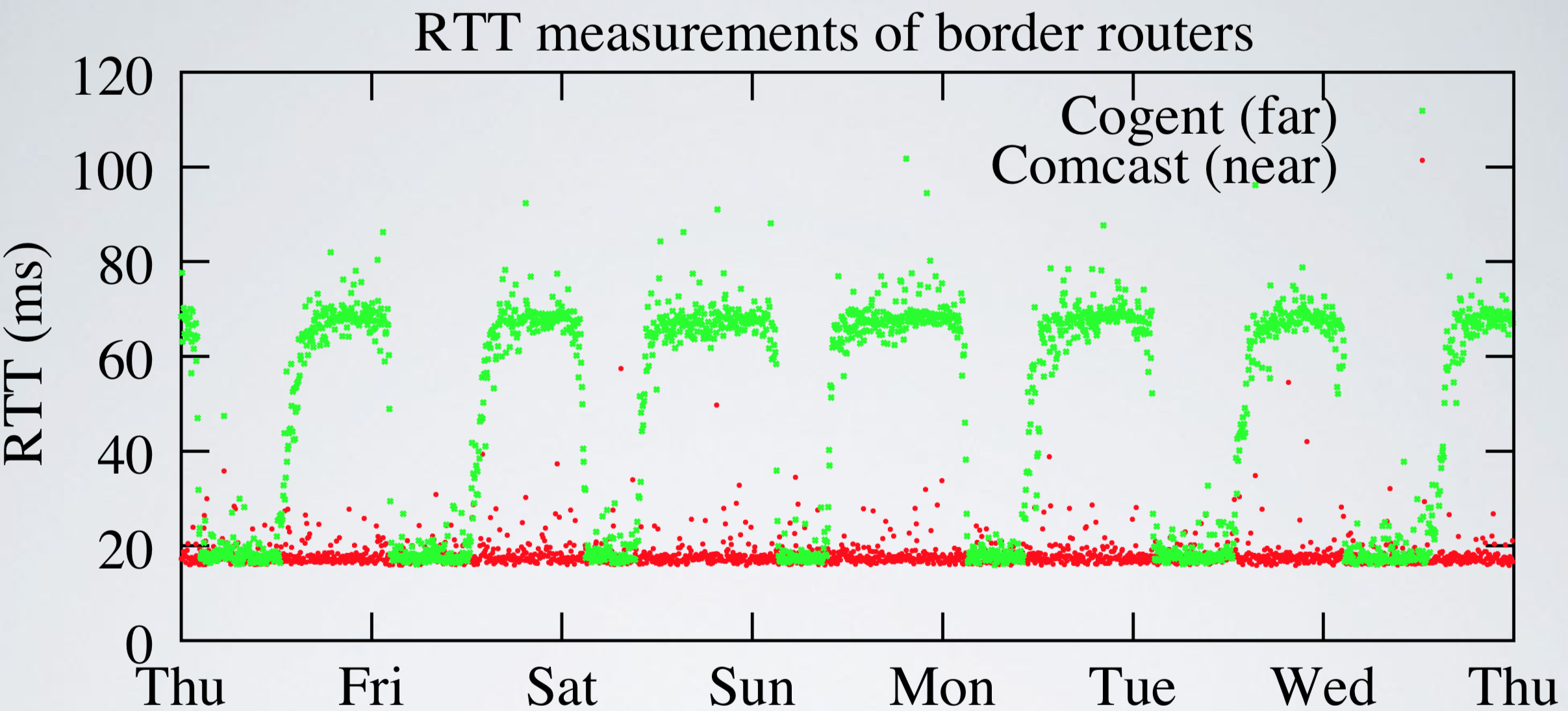}
    \captionof{figure}{Days of the week from 9/11/2017 to 16/11/2017 in New York city. It can be seen that congestion is higher during the weekend than weekdays. Monday, 13th November was a holiday for Veteran's Day. hence congestion is high on that day.}
    \label{fig2}
\end{center}

The time series in stored in InfluxDB which provides a SQL-like interface to query data and aggregates. The RTT is measured for each interconnecting link of near and far IP for a target for a particular ASN (Neighboring Autonomous System). The network service provider is determined by the \textit{monitor} tag. This has been in use from April, 2016 and a probe is sent every 5 minutes for every interconnecting link and the RTT is recorded \cite{ref1}. CAIDA is currently working on adding data prior to 2016 into the database.

\section{Correlations}

\subsection{Pearson Correlation Coefficient}

This coefficient measures the statistical association between two continuous variables based on their co-variance\cite{ref6}. As two time series are independent of each other, the pearson correlation coefficient method can be applied.

\begin{center}
    \centering
    \includegraphics[scale=0.23]{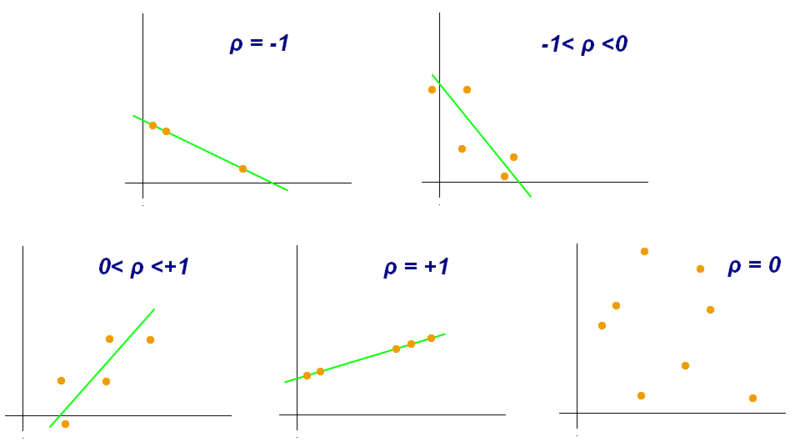}
    \captionof{figure}{Scatter diagrams with different coefficient values}
    \label{fig3}
\end{center}

The correlation $\rho$(X,Y) is given by:

\begin{center}
    \centering
    \includegraphics[scale=0.5]{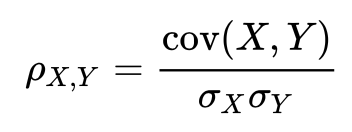}
    \label{fig4}
\end{center}

where cov is co-variance and $\sigma$(X) and $\sigma$(Y) are the standard deviations of X and Y respectively.\\
\\
Co-variance of X and Y is given wrt Expectation (E) by: 

\begin{center}
    \centering
    \includegraphics[scale=0.5]{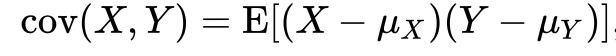}
    \label{fig5}
\end{center}

\subsection{Implementation}

We start by extracting all near-side time series for a monitor to a certain neighbor ASN, i.e., the \textit{ind} field is 0. An example would be all time series for monitor bed-us (Comcast) to Google (ASN 15169) with ind = 0. Similarly extract all the far-side time series for a certain monitor to a neighbor ASN, i.e., series where the ind = 1.\\
\\
A first analysis is to look at the correlations between different time series to the same far ASN. So, for the example of mon = bed-us and asn = 15169 (Comcast and Google), if you fix mon = bed-us and asn = 15169 and ind = 0, you get a number of time series where the \textit{target} field (and hence also the link field) is different. These are all the near-side interfaces corresponding to Comcast and Google seen from the bed-us monitor. Similarly, if you fix mon = bed-us and asn = 15169 and ind = 1 you will get all the time series for different target interfaces on the far-side.\\
\\
This tells us whether the time series representing different links from the same monitor behave similarly.

\begin{itemize}
    \item The Pearson Correlation Coefficient correlation between all pairs of near-side time series, i.e., ind = 0 is +0.059.
    \item The Pearson Correlation Coefficient correlation between all pairs of far-side time series, i.e., ind = 1 is +0.103.
\end{itemize}

Since both the coefficients are relatively low, we can see that there is no correlation between the near-side and far-side links for Comcast and Google. However, Verizon-Netflix had a higher correlation of 0.53 for target interfaces on near-side and 0.62 for target interfaces on far-side.\\
\\

Further, we look at correlations between the time series to different ASNs using the same ISP. For example, the time series for links from Comcast-Google as correlated with time series from Comcast to other ASNs such as Amazon, Netflix, TATA, etc. Network congestion on Comcast between Google and Netflix had low correlations, while Amazon and Netflix time series on Verizon had higher correlations.\\
\\
However, it must be kept in mind that a correlation does not imply causation. They simply imply a pattern in the congestion rates of the corresponding ISP and ASN. A simple increase in network capacity could suffice, or this could be a result of some ongoing disputes. Hence, these correlations have to analyzed further to get concrete results. Inferences from these correlations are being drawn by the policy team at CAIDA.

\section{Time-Series Forecasting}

Holt-Winters method is also known as Triple Exponential Smoothing due to the effect these methods have on a graph if you plot the values i.e. sharp lines become smoother \cite{ref3}. It uses the concept of weighted moving averages to predict the next value in the series. The most appropriate \textit{weight} is found through reducing error over multiple iterations.

\subsection{Single Exponential Smoothing}

Here, we consider the weighted average of all the data points and assign weights which reduce exponentially as we go back in time i.e. the most recent data points are given the most weightage in a recursive manner.\\
\\
For example, the weights could be $0.7^1$, $0.7^2$, $0.7^3$, $0.7^4$, $0.7^5$...\\
\\
While computing weighted average in this manner, the weights are normalized i.e. sum of weights is 1. However, here we see that this is not the case. To overcome this, we use the Poison equation:

\begin{center}
    \centering
    \includegraphics[scale=0.5]{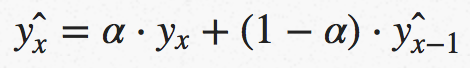}
    \label{fig6}
\end{center}

Here, the expected value is the sum of $\alpha*y_x$ and $(1-\alpha)$* \^{y}\textsubscript{x-1}. $\alpha$ can be thought of as the starting weight, previously 0.7. Thus, this becomes a weighted average with two weights $\alpha$ and $(1-\alpha)$ whose sum is 1. Because $(1-\alpha)$ is the weight of the x-1 instance, this gives the same effect as the original recursion and $(1-\alpha)$ is multiplied with itself with every recursion. Hence, it is called \textit{exponential} smoothing.\\
\\
$\alpha$ can also be called the \textit{decay rate} as higher the $\alpha$,  lower the weightage of the previous data points. $\alpha$ is then fitted using Mean Squared Error (MSE). 

\subsection{Double Exponential Smoothing}

This method is used to forecast a series of points instead of a single point as before. To do this, we require an additional factor \textit{b} called the \textit{trend} which looks at the ratio of the values current to previous point. The previously predicted value is termed as \textit{level} for the next computation.\\
\\
The equations are as followed:

\begin{center}
    \centering
    \includegraphics[scale=0.5]{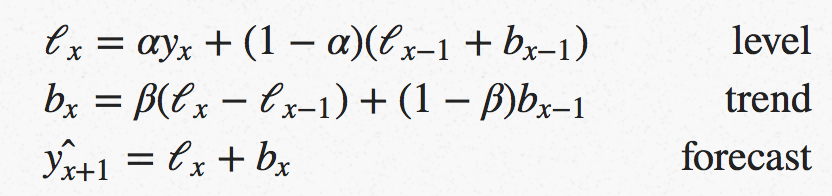}
    \label{fig7}
\end{center}

where $\beta$ is the \textit{trend factor}. The best fit is computed similarly to $\alpha$

\subsection{Triple Exponential Smoothing}

This method adds yet another parameter to the forecasting. If repetitive patterns exist at regular intervals in the series, the pattern is called a \textit{season} and the series is said to be seasonal \cite{ref3}. Every point in a season has a seasonal component which is a deviation from (level+trend). A series must be \textit{seasonal} to be forecasted using Holt-Winters method. From \ref{fig2}, we see that RTT measurement is seasonal.\\
\\
Smoothing is also applied across seasons in addition to level and trend. So, the 5\textsuperscript{th} point of the 7\textsuperscript{th} season would be smoothed with the 5\textsuperscript{th} point from the previous two seasons.

The equations follow these results:

\begin{center}
    \centering
    \includegraphics[scale=0.5]{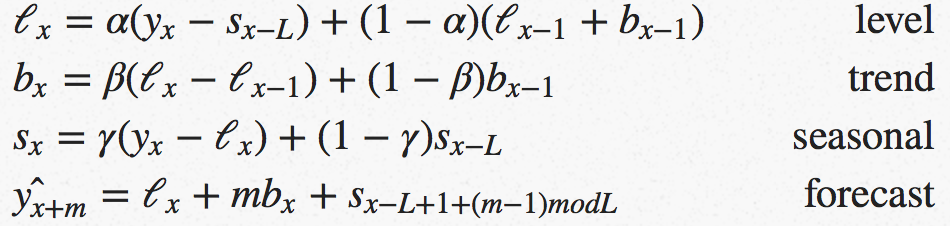}
    \label{fig8}
\end{center}

Here, $\gamma$ is seasonal smoothing factor, \textit{m} is the number of points we want to forecast, and \textit{L} is the season length.

\subsection{Results}

This method was applied to the Comcast-Netflix link and it was able to predict time series with an accuracy of 62\%. 

\begin{center}
    \centering
    \includegraphics[scale=0.41]{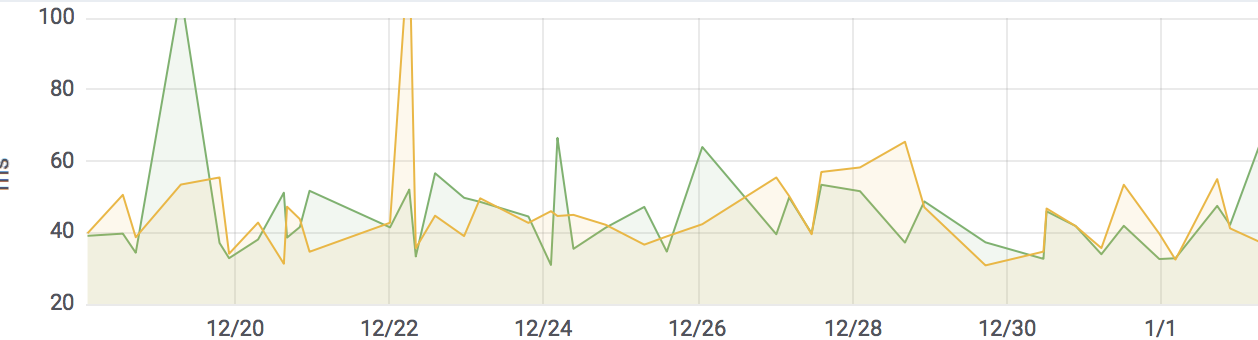}
    \captionof{figure}{Time series forecasting of Comcast-Netflix link from 20/12/17 to 1/1/18. Green line depicts actual RTT (ms) while yellow is predicted RTT (ms).}
    \label{fig9}
\end{center}

The smoothing factors took a huge amount of iterations to reach a desired accuracy, and hence trial and error had to be used. This was a factor in producing a relatively low accuracy.

\section{Challenges and Future Scope}

The internet is vast and dynamic, changing every second. Thus, new links are created and destroyed everyday. While this project takes a step further into the understanding of this network, many improvements and extensions can be made to allow for better analysis. \\
\\
Certain monitors may go offline at certain points in time which leads to a loss of data until they are turned back on. A method needs to be implemented to interpolate this data based on the time series available to us.\\
\\
Currently, the identification of all inter-domain links and their owners is not possible due to the complexity of the internet. Thus, accurately identifying all links requires tools which are currently not available to us.\\
\\
To overcome the performance bottleneck of computing the correlations between time series, as well as smoothing coefficients, the process can be parallelized on Comet i.e. the supercomputer at CAIDA. This improves the speed of computations and thus provides real-time results. This is currently being implemented.\\
\\
This time series forecasting information can be used to predict network congestion in different scenarios as certain policies are changed. This simulation would help in developing an all-rounded policy that governs service and content providers, to arrive at an informed conclusion on the net neutrality debate.

\end{document}